# NEUTRON MULTIPLICITY COUNTING DISTRIBUTION RECONSTRUCTION FROM MOMENTS USING MEIXNER POLYNOMIAL EXPANSION AND N-FORKED BRANCHING APPROXIMATIONS

**Philippe Humbert**

CEA, DAM, DIF, F-91297 Arpajon, France

philippe.humbert@cea.fr

**ABSTRACT**

Methods used to infer nuclear parameters from neutron count statistics fall into two categories depending on whether they use moments or count number probabilities. As probabilities are in general more difficult to calculate, we are interested here in the reconstruction of distributions from their first moments. For this, we explore two approaches. The first one relies on a generalization of the 2-forked branching correlation (quadratic) approximation used in the PMZBB and Poisson radical distributions, the second one is based on the expansion of the distribution on a basis of Meixner discrete orthogonal polynomials.

## I. INTRODUCTION

In a fissile system, neutrons originating from the same fission chain are time correlated. Therefore, the measurement of neutron time correlations is one of the major techniques used to detect and characterize the presence of nuclear materials[1,2]. This approach is used in many areas such as nuclear security, safeguards, criticality safety, or nuclear material inventory. The statistical analysis of the number of detected neutrons recorded per time interval is the basis of most of these techniques.

The majority of these methods are based on the first moments of the distribution, which are easy to interpret under a point model approximation where analytical expressions relate the moments to the system parameters. In general, the first three moments are used, namely, the single, double and triple counts in the Böhnel formalism[3] or, equivalently, the average count rate and the second and third Feynman correlated moment[4,5].

Alternatively, approaches based on the count number probability distribution have also been applied although the link to the parameters is more complicated because the probability of zero count is solution of a non-linear equation and the number of equations to be solved can become large when the count rate is high. However, the probability method can provide better results than the moments as shown by J. Verbeke[6] who used the count distribution for Bayesian parameters credible region estimation.

Probability methods were also used in the past thanks to certain approximations. Pàl[7], Mogil'ner and Zolotukhin[8], Bell[9] and Babala[10] derived the so-called PMZBB distribution, which is based upon the quadratic or two-forked branching approximation where the fission chains leading to the detection can be represented by binary trees. In the case of weak correlations, it was noted that the PMZBB distribution is close to the negative binomial distribution or Polya distribution[11], which is completely defined by its first two moments. On the other hand, the asymptotic form of the PMZBB distribution is the Radical Poisson distribution introduced by Nicolas Pacilio[12]. This distribution is also parametrized by its first two moments.

It seems therefore interesting to generalize these methods to reconstruct the counting distributions knowing their first moments. When the probability of zero detections is negligible and the average count is large, Moussa and Prinja[13] proposed to use the development on the generalized Laguerre polynomials that are associated with the continuous gamma distribution; they also study a maximum entropy method of reconstruction. Here we consider two alternative moment based reconstruction methods based on improvement of the Polya and Poisson radical approximations.

The negative binomial distribution is the measure associated to the Meixner discrete orthogonal polynomials[14], thus, we intend to improve the Polyà approximation with higher order moments by expanding the distribution on the Meixner polynomial basis. The other approach we have considered is the generalization of the radical Poisson distribution performed by taking into account higher order correlations, namely, the three and four forked branching correlations, leading to the cubic and quartic approximations. These approximated distributions are fully defined knowing the first three and four moments of the distribution.

In the following, we start by presenting the calculation of probabilities and moments of the reference distribution in the framework of a lumped model. Then, we study the generalization of the Poisson radical distribution by taking into account the third and fourth order branching correlations and we consider the improvement of the Polyà approximation using Meixner Polynomial expansions. Finally, we present numerical tests to verify and illustrate these two reconstruction techniques.

## II. REFERENCE MULTIPLICITY DISTRIBUTION CALCULATION

The reference distribution is calculated using a time dependent point model approximation with a compound Poisson external source where each source event has a probability to produce more than one neutron. In the point model, the phase space is reduced to one point; mono-kinetic neutrons are moving in an infinite homogeneous space. The interactions with the material media is limited to fission and capture. The capture event regroups all the non-fission events including leakage. The derivation of the equations for the detection number probability distribution is carried out using the adjoint Pàl[15] and Bell[16] methodology, which proceeds in two steps. The distribution induced by a single initial neutron is derived and used in a second phase to obtain the distribution in presence of an external source.

### II.A. Distribution Induced by a Single Neutron

*II.A.1. Probability Generating Function*

We consider the probability $p(n, T|i, t)$ to count $n$ neutrons at final time $T$ given that there was $i$ neutrons present at initial time $t$. We adopt the following simplified notations:

$$p_{n,i}(t) = p(n, T|i, t)$$

$$p_n(t) = p_{n,1}(t)$$

The probability equation is obtained from the balance of all possible events for one neutron in the elementary time interval $[t, t + dt]$ that lead to the counting of $n$ neutrons at final time $T$. The probability equation and final time condition are:

$$-\theta \frac{dp_n}{dt} + p_n = \frac{k}{v_1} \sum_{i=0}^{\infty} f_i p_{n,i} + \left(1 - \frac{k}{v_1}\right)\left((1 - \varepsilon_C)\delta_{n,1} + \varepsilon_C \delta_{n,0}\right) \quad (1)$$

$$p_n(T) = \delta_{n,0}$$

In equation (1) $\varepsilon_C$ is the capture detection efficiency, i.e. the average number of detections for one capture. The detector is open during a time gate $[0, T]$. Using the heaviside unit step function $H(t)$, the efficiency is:

$$\varepsilon_C(t) = \varepsilon_C\big(H(t) - H(t-T)\big) = \begin{cases} \varepsilon_C, & t \in [0,T] \\ 0, & t \notin [0,T] \end{cases}$$

Delayed neutrons production is neglected because the time gate width $T$ is short compared to the precursor decay time constant.

$f_i$ is the probability for one induced fission event to produce $i$ neutrons and $\nu_1$ is the average number of emitted neutrons.

$$\nu_1 = \sum_{i=0}^{\infty} i f_i$$

More generally $\nu_i$ is the binomial moment of order $n$ of the induced fission multiplicity.

$$\nu_i = \sum_{n=i}^{\infty} \binom{n}{i} f_i$$

$k$ is the multiplication factor, $\dfrac{k}{\nu_1}$ is the fission probability and $\theta$ is the neutron average lifetime.

In order to deal with the $p_{n,i}$ term and to derive the binomial moment equations we introduce the probability generating function (PGF).

$$g(x) = \sum_{n=0}^{\infty} x^n p_n \quad \text{with} \quad x \in [0., 1.]$$

We will use the following PGF property:

$$\big(g(x)\big)^i = \sum_{n=0}^{\infty} x^n p_{n,i}$$

The binomial moments of the counting distribution are defined by:

$$m_n = \sum_{k=n}^{\infty} \binom{k}{n} p_k \qquad m_{n,i} = \sum_{k=n}^{\infty} \binom{k}{n} p_{k,i}$$

The probabilities and binomial moments can be obtained by derivation of the PGF.

$$p_n = \frac{1}{n!}\left(\frac{\partial^n g}{\partial x^n}\right)_{x=0} \qquad p_{n,i} = \frac{1}{n!}\left(\frac{\partial^n g^i}{\partial x^n}\right)_{x=0}$$

$$m_n = \frac{1}{n!}\left(\frac{\partial^n g}{\partial x^n}\right)_{x=1} \qquad m_{n,i} = \frac{1}{n!}\left(\frac{\partial^n g^i}{\partial x^n}\right)_{x=1}$$

Applying the PDF definition in (1) we get the PGF equation.

$$-\theta\frac{\partial g}{\partial t} + g(x,t) = \frac{k}{v_1}\sum_{i=0}^{I} f_i g^i + \left(1 - \frac{k}{v_1}\right)(1 - \varepsilon_C(1-x)) \qquad (2)$$

$$g(x,T) = 1$$

*II.A.2. Detection Number probabilities*

The probability equations are obtained by the successive derivatives of equation (2). The probability of zero detection is $p_0 = g(0)$.

$$-\theta\frac{dp_0}{dt} + p_0 - \frac{k}{v_1}\sum_{i=0}^{I} f_i p_0^i = \left(1 - \frac{k}{v_1}\right)(1 - \varepsilon_C) \qquad p_0(T) = 1$$

The probability of one detection is $p_1 = \left(\frac{\partial g}{\partial x}\right)_{x=0}$.

$$-\theta\frac{dp_1}{dt} + \left(1 - \frac{k}{v_1}\sum_{i=0}^{I} if_i p_0^{i-1}\right) p_1 = \left(1 - \frac{k}{v_1}\right)\varepsilon_C \qquad p_1(T) = 0$$

The subsequent probabilities for *n* greater than one are:

$$-\theta\frac{dp_n}{dt} + p_n = \frac{k}{v_1}\sum_{i=0}^{I} f_i p_{n,i} \qquad p_n(T) = \delta_{n,0}$$

In order to recursively calculate $p_{n,i}$ we use the Leibnitz recurrence formula for the derivatives of compound functions.

$$\left(\frac{\partial^n g^i}{\partial x^n}\right) = \left(\frac{\partial^n g^{i-1} g}{\partial x^n}\right) = \sum_{j=0}^{n} \binom{n}{j}\left(\frac{\partial^{n-j} g^{i-1}}{\partial x^{n-j}}\right)\left(\frac{\partial^k g}{\partial x^j}\right)$$

From this, we deduce that:

$$p_{n,i} = \sum_{j=0}^{n} p_{n-j,i-1} p_j \qquad (3)$$

The probability $p_{n,i}$ can be decomposed into two parts, the first one depends on $p_0$ and $p_n$, the second one, $u_{n,i}$, depends on the previously calculated probabilities $p_{n-1}, p_{n-1}, \cdots, p_0$

$$p_{n,i} = ip_0^{i-1} p_n + u_{n,i}$$

The probabilities are then calculated by solving for increasing *n* the following equation,

$$-\theta\frac{dp_n}{dt} + \left(1 - \frac{k}{v_1}\sum_{i=0}^{I} if_i p_0^{i-1}\right) p_n = \frac{k}{v_1}\sum_{i=2}^{I} f_i u_{n,i} \qquad p_n(T) = 0$$

The term $u_{n,i}$ is calculated using the following relation that comes from equation (3):

$$u_{n,1} = 0 \qquad u_{n,i>1} = u_{n,i-1}p_0 + \sum_{j=1}^{n-1} p_{n-j,i-1}p_j$$

For the numerical solution of these equations, the time derivatives are discretized using an implicit scheme and the nonlinear equation of $p_0$ is solved with the Newton method.

### II.A.3. Binomial Moments

The binomial moments are also the derivatives of the PGF. They are therefore solution of equations similar to those of the probabilities except for the moment of order zero that is just equal to one. Therefore, by analogy, we have:

$$m_0 = 1$$
$$-\theta \frac{dm_1}{dt} + (1-k)m_1 = \left(1 - \frac{k}{\nu_1}\right)\varepsilon_C \qquad m_1(T) = 0$$

For *n* greater than one, the equation is:

$$-\theta \frac{dm_n}{dt} + m_n = \frac{k}{\nu_1} \sum_{i=0}^{I} f_i m_{n,i} \qquad m_n(T) = \delta_{n,0}$$

Similar to the relation (3) we have:

$$m_{n,i} = \sum_{j=0}^{n} m_{n-j,i-1} m_j$$

The moment $m_{n,i}$ is decomposed into two parts, the first one depends on $m_n$, the second one, $w_{n,i}$, depends on the previous calculated moments $m_{n-1}, m_{n-2}, \cdots, m_1$.

$$m_{n,i} = i m_n + w_{n,i}$$

The binomial moments are then calculated by solving for increasing *n* the following equation:

$$-\theta \frac{dm_n}{dt} + (1-k)m_n = \frac{k}{\nu_1} \sum_{i=0}^{I} f_i w_{n,i} \qquad m_n(T) = 0$$

The term $w_{n,i}$ is calculated with:

$$w_{n,1} = 0 \qquad w_{n,i>1} = w_{n,i-1} + \sum_{j=1}^{n-1} m_{n-j,i-1} m_j$$

### II.A.4. Distribution Induced by a Single Initial Source Event

The external source can produce more than one neutron by source event, this is the case for the spontaneous fission source. We use the following notations:

$\hat{f}_i$ is the probability that one source event emits *i* neutrons.

$\hat{p}_n(t)$ is the probability to count $n$ neutrons at final time given that there was one source event at initial time $t$ and $\hat{m}_n$ is the corresponding binomial moment.

$$\hat{p}_n = \sum_{i=0}^{\infty} \hat{f}_i p_{n,i} \qquad \hat{m}_n = \sum_{i=0}^{\infty} \hat{f}_i m_{n,i}$$

The associated PGF is:

$$\hat{g}(x) = \sum_{n=0}^{\infty} x^n \hat{p}_n = \sum_{i=0}^{\infty} \hat{f}_i \big(g(x)\big)^i$$

**II.B. Distribution Induced by an External Source without Initial Neutrons**

*II.B.1. Probability Distribution*

We consider the probability $P_S(n, T | i, t)$ to count $n$ neutrons at final time $T$ in the presence of an external source given that there was $i$ neutrons present at initial time $t$. We adopt the following simplified notations:

$$P_{n,i}(t) = P_S(n, T | i, t)$$

$$P_n(t) = P_{n,0}(t)$$

The probability equation is obtained from the balance of all possible source events in the elementary time interval $[t, t + dt]$ that lead to the counting of $n$ neutrons at final time $T$. Noting that the source intensity (number of source events per time unit) is $S$, The probability equation and final time condition are:

$$-\frac{dP_n}{dt} = S\left(\sum_{i=0}^{\infty} \hat{f}_i P_{n,i} - P_n\right) \qquad P_n(T) = \delta_{n,0} \qquad (4)$$

The PGF associated to the probability $P_n$ is noted $G(x)$. We apply the PGF definition to equation (4) so that we have:

$$\frac{\partial G}{\partial t} = SG(1 - \hat{g}) \qquad G(x, T) = 1$$

Hence $G(x,t)$ is the PGF of a compound Poisson distribution.

$$G(x, t) = \exp\left(-\int_t^T S(\tau)\big(1 - \hat{g}(x, \tau)\big) dt\right)$$

The probability for zero detection is:

$$P_0(t) = G(0, t) = \exp\left(-\int_t^T S(\tau)\big(1 - \hat{p}_0(\tau)\big) dt\right)$$

By definition, the cumulant generating function is:

$$K(x,t) = \text{Log}(G(x,t)) = -\int_t^T S(t)(1-\hat{g}(x,t))dt$$

The parameters of the compound Poisson distribution are then:

$$\Lambda_n(t,T) = \frac{1}{n!}\left(\frac{d^n K}{dx^n}\right)_{x=0} = \int_t^T S(\tau)\hat{p}_n(\tau)d\tau$$

Thus, we can apply the Panjer's recurrence[17] for compound Poisson distribution to calculate the probabilities $P_n$.

$$nP_n = \sum_{k=1}^{n} k\Lambda_k P_{n-k}$$

*II.B.2. Binomial Moments, Cumulants and Feynman relative Moments*

The binomial cumulants are given by the derivatives of the cumulant generating function.

$$\Gamma_n(t,T) = \frac{1}{n!}\left(\frac{d^n K}{dx^n}\right)_{x=1} = \int_t^T S(t)\widehat{m}_n(t)dt$$

The binomial moment are successively calculated using the Panjer's relation:

$$nM_n = \sum_{k=1}^{n} k\Gamma_k M_{n-k}$$

When the source is constant and the system is in a stationary state, we have:

$$\Gamma_n(T) = \Gamma_n(-\infty, T) = \int_{-\infty}^{T} S\widehat{m}_n(t)dt$$

The Feynman moments are defined as:

$$Y_n(T) = \frac{n!\,\Gamma_n(T)}{\Gamma_1(T)}$$

In particular, the second order Feynman moment is the function originally defined by R. Feynman as the excess of reduced variance:

$$Y_2 = \frac{2\Gamma_2}{\Gamma_1} = \frac{V}{M_1} - 1$$

### III. N-FORKED BRANCHING APPROXIMATIONS

We are looking for an approximated distribution that can be calculated using only the first moments of the reference. For this purpose, we make several approximations. The count number distribution induced by one initial neutron is independent of time (stationarity). The phase space is reduced to one point (point model). The external source produces uncorrelated neutrons (Poisson source).

In the fission chain contributing to the detection each fission gives rise to N useful sub-chains at most (N-forked branching approximation).

### III.A. One Initial Neutron Probability Generating Function Equation

From equation (2), the stationary equation for the PGF $g(x)$ writes:

$$g(x) = \frac{k}{\nu_1} \sum_{i=0}^{\infty} f_i g^i + \left(1 - \frac{k}{\nu_1}\right)(1 - \varepsilon_C(1-x)) \tag{5}$$

In the N-forked approximation, the binomial moments of the fission multiplicity or order greater than N are neglected.

$$\nu_i = 0 \quad \text{for} \quad i > N$$

In order to obtain an equation which depends on the binomial moments $\nu_i$ we consider the following function:

$$h(x) = 1 - g(x)$$

We replace $g$ by $1-h$ in equation (5).

$$(1-k)h(x) = \frac{k}{\nu_1} \sum_{i=2}^{N} \nu_i(-h)^i + \varepsilon_C \left(1 - \frac{k}{\nu_1}\right)(1-x) \tag{6}$$

We introduce some new notations. The fission detection efficiency $\varepsilon_F$ is the average number of detections for one induced fission. It is related to the capture detection efficiency $\varepsilon_C$ by:

$$\frac{k}{\nu_1} \varepsilon_F = \left(1 - \frac{k}{\nu_1}\right) \varepsilon_C$$

In addition, we replace the multiplication coefficient by the reactivity $\rho$,

$$\rho = \frac{k-1}{k}$$

With these notations, equation (6) becomes:

$$h(x) = -\frac{1}{-\rho \nu_1} \sum_{i=2}^{N} (-1)^i \nu_i h^i + \frac{\varepsilon_F}{-\rho \nu_1}(1-x) \tag{7}$$

It is now appropriate to express the coefficients of this algebraic equation in terms of the one initial neutron binomial moments.

$$m_n = -\frac{1}{n!} \left(\frac{d^n h}{dx^n}\right)_{x=1}$$

The first moment is:

$$m_1 = -\left(\frac{dh}{dx}\right)_{x=1} = \frac{\varepsilon_F}{-\rho\nu_1}$$

The moments of higher order ($n > 1$) are:

$$m_n = \sum_{i=2}^{n} \mu_{n,i}$$

$$\mu_{n,i} = (-1)^i \frac{\nu_i}{-\rho\nu_1} \frac{1}{n!} \left(\frac{d^n h^i}{dx^n}\right)_{x=1} \qquad (8)$$

In particular, we have:

$$\mu_{i,i} = (-1)^i \frac{\nu_i}{-\rho\nu_1} \frac{1}{i!} \left(\frac{d^i h^i}{dx^i}\right)_{x=1} = \frac{\nu_i}{-\rho\nu_1} m_1^i$$

It is worthwhile to change the unknown function $h(x)$:

$$b(x) = \frac{h(x)}{m_1}$$

The polynomial equation (7) becomes:

$$\sum_{i=2}^{\infty} (-1)^i \frac{\mu_{i,i}}{m_1} b^i + b + x - 1 = 0$$

The problem is to obtain an expression in terms of the moments of the detection number distribution induced by a neutron source. For this purpose, we consider the binomial cumulants and the Feynman relative moments.

In case of an uncorrelated Poisson neutron source the binomial cumulants are given by:

$$\Gamma_n = ST m_n$$

The Feynman moments are a ratio of cumulants, and then a ratio of without source binomial moments.

$$Y_n = \frac{n! \Gamma_n}{\Gamma_1} = \frac{n! m_n}{m_1}$$

The Feynman moments can be written as a sum over partial Feynman moments which correspond to different branching configurations (cf. figure 1).

$$Y_{n,i} = n! \frac{\mu_{n,i}}{m_1} \qquad (9)$$

$$Y_n = Y_{n,n} + \sum_{i=2}^{n-1} Y_{n,i} \qquad (10)$$

In the two-forked branching case we have

$$Y_{2,2} = Y_2$$

The $b(x)$ equation is then:

$$\sum_{i=2}^{\infty}(-1)^i \frac{Y_{i,i}}{i!} b^i + b + x - 1 = 0$$

The with source PGF is:

$$G(x,T) = \exp(-M_1 b(x))$$

### III.B. Two-Forked Branching (Quadratic) Approximation

In the two-forked branching approximation only the first two moments of the fission multiplicity distribution are considered.

$$\nu_i = 0 \quad \text{for} \quad i > 2$$

Consequently, we have a second-degree polynomial equation for $b(x)$. The 2-forked approximation is a quadratic approximation.

$$\frac{Y_2}{2} b^2 + b + x - 1 = 0$$

The positive solution is:

$$b(x) = \frac{1}{Y_2}\left(\sqrt{1 + 2Y_2(1-x)} - 1\right)$$

The PGF has a closed form expression.

$$G(x) = \exp\left(-\frac{M_1}{Y_2}\left(\sqrt{1 + 2Y_2(1-x)} - 1\right)\right) \tag{11}$$

This is the stationary PMZBB also called Poisson radical generating function expression.

The second order Feynman moment expression in terms of binomial moments is obtained from Panjer's recurrence.

$$\Gamma_1 = M_1 \qquad 2\Gamma_2 = 2M_2 - M_1^2$$

$$Y_2 = \frac{2M_2}{M_1} - M_1$$

In the quadratic approximation, the detection number probability distribution is completely defined given the first two moments of the exact distribution.

### III.C. Three-Forked Branching (cubic) Approximation

In the three-forked branching approximation only the first three moments of the fission multiplicity distribution are considered.

$$\nu_i = 0 \quad \text{for} \quad i > 3$$

Consequently, we have a third-degree polynomial equation for $b(x)$. The 3-forked approximation is a cubic approximation.

$$-\frac{Y_{3,3}}{3!}b^3 + \frac{Y_2}{2}b^2 + b + x - 1 = 0$$

An analytical solution can be obtained using the Cardano method or alternatively solved numerically using the Newton method

We derive the expression of the partial Feynman moment $Y_{3,3}$ in terms of $Y_2$ and $Y_3$. From equation (10) we have:

$$Y_{3,3} = Y_3 - Y_{3,2}$$

According to equation (9) we have:

$$Y_{3,2} = 3!\frac{\mu_{3,2}}{m_1}$$

From equation (8) we have:

$$\mu_{3,2} = \frac{\nu_2}{-\rho\nu}\frac{1}{3!}\left(\frac{d^3 h^2}{dx^3}\right)_{x=1}$$

After performing the derivatives we obtain:

$$\mu_{3,2} = \frac{Y_2^2}{2}m_1$$

$$Y_{3,2} = 3Y_2^2$$

Finally we obtain the expression of $Y_{3,3}$:

$$Y_{3,3} = Y_3 - 3Y_2^2$$

This result can be interpreted in terms of branching configurations as illustrated in figure 1. More precisely, in order to enumerate all the branching configurations, the detections are numbered and detected neutrons that belong to sibling fission chains are bracketed.

$$Y_{3,3} : (1,2,3)$$

$$3Y_2^2 : \big(1,(2,3)\big)\big(2,(1,3)\big)\big(3,(1,2)\big)$$

Thus, we obtain:

$$Y_3 = Y_{3,3} + 3Y_2^2$$

The third order Feynman moment expression in terms of moments is obtained using the Panjer's recurrence.

$$Y_3 = 6\left(\frac{M_3}{M_1} - M_2\right) + 2M_1^2$$

Thus, in the cubic approximation, the detection number probability distribution is completely defined given the first three moments of the exact distribution.

### III.C. Four-Forked Branching (quartic) Approximation

In the four-forked branching approximation only the first four moments of the fission multiplicity distribution are considered:

$$\nu_i = 0 \quad \text{for} \quad i > 4$$

Consequently, we have a fourth-degree polynomial equation for $b(x)$. The 4-forked approximation is a quartic approximation.

$$\frac{Y_{4,4}}{4!}b^4 - \frac{Y_{3,3}}{3!}b^3 + \frac{Y_2}{2}b^2 + b + x - 1 = 0$$

This equation can be solved numerically using the Newton method.

We derive the expression of the partial Feynman moment $Y_{4,4}$ in terms of $Y_2$, $Y_3$ and $Y_4$. From equation (10) we have:

$$Y_{4,4} = Y_4 - Y_{4,3} - Y_{4,2}$$

According to equation (9) we have:

$$Y_{4,3} = 4!\frac{\mu_{4,3}}{m_1}$$

From equation (8) we have:

$$\mu_{4,3} = \frac{\nu_3}{-\rho\nu}\frac{1}{4!}\left(\frac{d^4 h^3}{dx^4}\right)_{x=1}$$

After performing the derivatives we obtain:

$$\mu_{4,3} = 3\mu_{3,3}\frac{Y_2}{2}$$

$$Y_{4,3} = 6Y_{3,3}Y_2$$

From equation (8) we have:

$$\mu_{4,2} = \frac{\nu_2}{-\rho\nu}\frac{1}{4!}\left(\frac{d^4 h^2}{dx^4}\right)_{x=1}$$

After performing the derivatives we obtain:

$$\mu_{4,2} = 1 = 5\left(\frac{Y_2}{2}\right)^3 m_1 + 2\left(\frac{Y_2}{2}\right)\mu_{3,3}$$

$$Y_{4,2} = 15Y_2^3 + 4Y_2Y_{3,3}$$

So the expression of $Y_{4,4}$ is:

$$Y_{4,4} = Y_4 - 10Y_2Y_{3,3} - 15Y_2^3$$

Using the expression of $Y_{3,3}$ we finally have:

$$Y_{4,4} = Y_4 - 10Y_2Y_3 + 15Y_2^3$$

This result can be interpreted in terms of branching configurations as illustrated in figure 1. More precisely, in order to enumerate all the branching configurations, the detections are numbered and detected neutrons that belong to sibling fission chains are bracketed.

$Y_{4,4}$ : (1,2,3,4)

$6Y_{3,3}Y_2$ : $(1,2,(3,4))$ $(1,3,(2,4))$ $(1,4,(2,3))$ $(2,3,(1,4))$ $(2,4,(1,3))$ $(3,4,(1,2))$

$4Y_{3,3}Y_2$ : $(1,(2,3,4))$ $(2,(1,3,4))$ $(3,(1,2,4))$ $(4,(1,3,4))$

$3Y_2^3$ : $((1,2),(3,4))$ $((1,3),(2,4))$ $((1,4),(2,3))$

$12Y_2^3$ : $\left(1,(2,(3,4))\right)$ $\left(1,(3,(2,4))\right)$ $\left(1,(4,(2,3))\right)$
$\left(2,(1,(3,4))\right)$ $\left(2,(3,(1,4))\right)$ $\left(2,(4,(1,3))\right)$
$\left(3,(1,(2,4))\right)$ $\left(3,(2,(1,4))\right)$ $\left(3,(4,(1,2))\right)$
$\left(4,(1,(2,3))\right)$ $\left(4,(2,(1,3))\right)$ $\left(4,(3,(1,2))\right)$

Thus, we have:

$$Y_4 = Y_{4,4} + 10Y_2Y_{3,3} + 15Y_2^3$$

The fourth order Feynman moment expression in terms of moments is obtained using the Panjer's recurrence.

$$Y_4 = 24\frac{M_4}{M_1} - 6M_3 - 6Y_2M_2 - 3Y_3M_1$$

Therefore, in the quartic approximation, the detection number probability distribution is completely defined given the first four moments of the exact distribution.

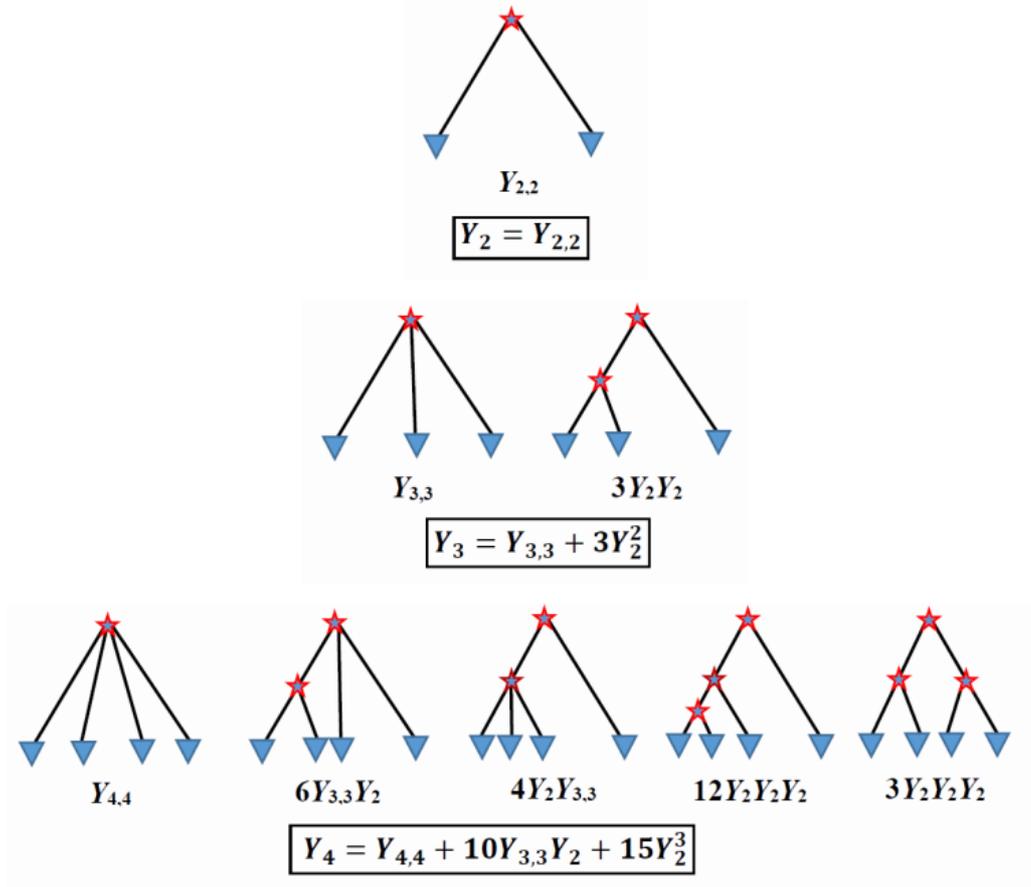

**Figure 1.** Fission branching configurations for two, three and four correlated detections used for the interpretation of the terms that appear in $Y_2$, $Y_3$ and $Y_4$ expressions. The stars are induced fissions, the triangles are the detections and the black lines represent the induced-fission chains.

### III.D. Probability Distributions Calculation

We have derived the expression of the PGF in terms of binomial moments. We have to invert the generating function in order to get the probability distribution.

The function $b(x)$ is solution of a polynomial equation:

$$\sum_{i=0}^{N} a_i \bigl(b(x)\bigr)^i = 0 \qquad (12)$$

The $a_i$ coefficients are:

$$a_0 = x - 1 \qquad a_1 = 1 \qquad a_{i>1} = (-1)^i \frac{Y_{i,i}}{i!}$$

We will use the following simplified notation for the derivatives of function $b(x)$ at $x=0$:

$$b_n = \frac{1}{n!}\left(\frac{d^n b}{dx^n}\right)_{x=0}$$

$b_0$ is solution of the polynomial equation:

$$\sum_{i=1}^{N} a_i b_0^i = 1$$

Analytical solutions exist for the equations of degree two and three, the fourth degree equation is solved by Newton's method.

In order to calculate $b_1$ we take the derivative of equation (12) in $x = 0$, we obtain:

$$b_1 = \frac{1}{\sum_{i=1}^{N} i a_i b_0^{i-1}}$$

To calculate the $b_n$ terms we use a methodology similar to the one presented in the previous chapter for the reference distribution calculation. We take the derivatives of equation (11) at $x = 0$ and we use the notation:

$$b_{n,i} = \frac{1}{n!}\left(\frac{d^n b^i}{dx^n}\right)_{x=0} \quad b_n = b_{n,1}$$

We obtain the $b_{n,i}$ equation:

$$\sum_{i=1}^{N} a_i b_{n,i} = 0$$

Using the Leibnitz recurrence for the derivatives of the power function, we have:

$$b_{n,i} = \sum_{k=0}^{n} b_{n-k,i-1} b_k$$

We note $\alpha_{n,i}$ the part of $b_{n,i}$ which depends on $b_{n-1}, b_{n-2},\ldots,b_1,b_0$

$$b_{n,i} = i b_0^{i-1} b_n + \alpha_{n,i}$$

$\alpha_{n,i}$ terms are calculated using:

$$\alpha_{n,1} = 0 \quad \alpha_{n,i>1} = \alpha_{n,i-1} b_0 + \sum_{k=1}^{n-1} b_{n-k,i-1} b_k$$

Finally, the $b_n$ are given by:

$$b_n = \frac{\sum_{i=1}^{N} a_i \alpha_{n,i}}{\sum_{i=1}^{N} i a_i b_0^{i-1}}$$

In the presence of an external source, the count number probabilities are successively calculated using the Panjer's recurrence:

$$P_0 = G(0) = \exp(-M_1 b_0)$$

$$n P_{n>0} = M_1 \sum_{i=0}^{n} b_i P_{n-i}$$

**IV. MEIXNER POLYNOMIAL EXPANSION**

When the multiplication of the measured fissile assembly is from low to moderate, it has been recognized by Mogilner et al. and Pacilio that the counting distribution can be approximated by a Polyà distribution. The first two moments of the Polyà distribution are free parameters that can be chosen to be those of the reference counting distribution. Moreover, the Polyà distribution is associated to the Meixner discrete orthogonal polynomials. A better reconstruction, based on the moments of the reference distribution will be obtained with a truncated expansion of the reference distribution on Meixner polynomials.

**IV.A. Polyà and Poisson Approximations**

When the neutrons are weakly correlated ($Y_2 \ll 1$), we have the following second order approximations:

$$\sqrt{1 + 2Y_2(1-x)} - 1 \cong \text{Log}(1 + (1-x)Y_2) \cong 1 + Y_2(1-x) - \frac{1}{2}\big(Y_2(1-x)\big)^2$$

Consequently, the Poisson radical PGF (cf. equation (11)) can be approximated by:

$$G(x,T) \cong \exp\left(-\frac{M_1}{Y_2} \text{Log}(1 + (1-x)Y_2)\right) = \left(\frac{1}{1 + (1-x)Y_2}\right)^{\frac{M_1}{Y_2}}$$

Here, we recognize the PGF of the Polyà distribution.

When the neutrons are uncorrelated ($Y_2 = 0$), $G(x,T)$ is the PGF of a Poisson distribution parametrized by the first moment only.

$$G(x) = \exp(-M_1(1-x))$$

$$P_n = \phi_{Poisson}^{(M_1)}(n) = \frac{M_1^n}{n!} \exp(-M_1)$$

**IV.B. Polyà Probability distribution**

The negative binomial distribution or Polyà's law is a discrete distribution which depends on two parameters $\alpha \in \mathbb{R}^+$ and $q \in [0., 1.]$. The probability mass function is:

$$\phi_{Polya}^{(\alpha,q)}(n) = \frac{\Gamma(n+\alpha)}{n!\,\Gamma(\alpha)} q^n (1-q)^\alpha$$

Here, $\Gamma()$ is the gamma function, which generalizes the factorial, for $n$ integer, $n! = \Gamma(n+1)$.

The mean and the variance are:

$$E(n) = \bar{n} = \frac{\alpha q}{1-q} \qquad V(n) = \sigma^2 = \frac{\bar{n}}{1-q}$$

The distribution is over-dispersed since the variance is $1/(1-q)$ greater than the mean. The Polyà distribution is a special case of compound Poisson distribution. It describes a branching process with a primary Poisson process with a secondary logarithmic distribution.

**IV.C. Meixner Discrete Orthogonal Polynomials**

The neutron counting distribution is close to the Polyà distribution, which is the measure associated to the Meixner discrete orthogonal polynomials. Thus, the Polyà approximation of the counting distribution can be improved by expanding the distribution on a Meixner polynomial basis. The coefficient of the expansion are calculated using the moments of the reference counting distribution.

The Meixner polynomials are expressed as:

$$\mathcal{M}_k^{(\alpha,q)}(n) = \sum_{i=0}^{k} (-1)^i \frac{k!}{(k-i)!} \frac{\Gamma(\alpha)}{\Gamma(\alpha+i)} \left(\frac{1-q}{q}\right)^i \binom{n}{i}$$

The orthogonality relation is:

$$\sum_{n=0}^{\infty} \mathcal{M}_k^{(\alpha,q)}(n) \mathcal{M}_l^{(\alpha,q)}(n) \phi_{Polya}^{(\alpha,q)}(n) = h_k \delta_{k,l}$$

The normalization coefficient is:

$$h_k = \frac{k!}{q^k} \frac{\Gamma(\alpha)}{\Gamma(\alpha+k)}$$

Meixner polynomials can be successively calculated using a three-points recurrence relation:

$$\mathcal{M}_0^{(\alpha,q)}(n) = 1$$

$$\mathcal{M}_1^{(\alpha,q)}(n) = 1 - \frac{1-q}{\alpha q} n$$

$$\mathcal{M}_{k+1}^{(\alpha,q)}(n) = \frac{1}{q(\alpha+k)} \left\{ [(1+q)k - (1-q)n + \alpha q] \mathcal{M}_k^{(\alpha,q)}(n) - k \mathcal{M}_{k-1}^{(\alpha,q)}(n) \right\}$$

**IV.D. Meixner Polynomial Approximation of the Counting Distribution**

The neutron counting distribution is expanded on the Meixner polynomial basis.

$$P_n = \phi_{Polya}^{(\alpha,q)}(n) \left( 1 + \sum_{k=1}^{\infty} c_k \mathcal{M}_k^{(\alpha,q)}(n) \right)$$

The coefficients $c_k$ are:

$$c_k = \frac{1}{h_k} \sum_{n=0}^{\infty} \mathcal{M}_k^{(\alpha,q)}(n) P_n$$

$$c_k = \frac{\Gamma(\alpha+k)}{\Gamma(\alpha)} \frac{q^k}{k!} \sum_{i=0}^{k} (-1)^i \frac{k!}{(k-i)!} \frac{\Gamma(\alpha)}{\Gamma(\alpha+i)} \left(\frac{1-q}{q}\right)^i M_i$$

Here, $M_i$ is the order $i$ binomial moments of the neutron counting distribution. The parameters $\alpha$ and $q$ are chosen in such a way that the coefficients $c_1$ and $c_2$ are null.

$$q = \frac{2M_2 - M_1^2}{2M_2 - M_1^2 + M_2} \qquad \alpha = \frac{M_1^2}{2M_2 - M_1^2}$$

An approximation of the neutron counting distribution is obtained by truncating the polynomial expansion:

$$P_n \approx \phi_{Meixner}^{[K]}(n) = \phi_{Polya}^{(\alpha,q)}(n) \left(1 + \sum_{k=3}^{K} c_k(M_1, M_2, \cdots, M_k) \mathcal{M}_k^{(\alpha,q)}(n)\right)$$

In particular, the order 2 approximation is the Polyà distribution which has the same two first moments as the neutron counting distribution.

$$\phi_{Meixner}^{[2]}(n) = \phi_{Polya}^{(\alpha,q)}(n)$$

## V. NUMERICAL APPLICATIONS

We present a series of test cases to verify and quantify the quality of the reconstructed distributions obtained with the different studied approximations. Namely, we apply on the one hand the Polyà approximation with its improvement using Meixner polynomial expansions and in the other hand, the quadratic (Poisson radical, 2-forked branching), cubic (3-forked branching) and quartic (4-forked branching) approximations. In general, we consider distributions reconstructed using the first two, three or four moments. In the different cases, the reference count distribution and its moments are calculated using the recursive method presented in chapter two. The parameters common to all test cases are given in table I.

**TABLE I**. Test Cases common parameters.

| Multiplication coefficient | $k$ | 0.90 |
|---|---|---|
| Prompt neutron decay constant | $\alpha$ (ms$^{-1}$) | 1. |
| Detection time gate width | $T$ (ms) | 10. |
| Average induced fission multiplicity | $\nu_1$ | 2.551 |
| Standard deviation | $\sigma$ | 1.08 |

Two types of neutron sources are studied. It is either an ($\alpha$,n) type source where each source event produces a single neutron or a spontaneous fission type of source where each fission has a probability distribution to emit zero, one or more neutrons. The spontaneous and induced fission multiplicities are Terrell truncated Gaussian distributions[18] defined by the mean and the standard deviation. We also study two levels of detection efficiency and the corresponding source levels are adjusted in order to achieve the same count rate. We thus define four test cases whose specific parameters are given in table II.

**TABLE II.** Tests cases specific parameters.

|        | Detection efficiency | Source intensity | Source type         |                                          | $Y_2$   |
|--------|----------------------|------------------|---------------------|------------------------------------------|---------|
| Case 1 | 0.01                 | 10 n/ms          | (α,n)               | $\hat{f}_n = \delta_{n,1}$               | 1.0668  |
| Case 2 | 0.01                 | 10 n/ms          | Spontaneous fission | $\nu_{S1} = 3.76$ $\sigma = 1.245$       | 1.2528  |
| Case 3 | 0.1                  | 1 n/ms           | (α,n)               | $\hat{f}_n = \delta_{n,1}$               | 10.668  |
| Case 4 | 0.1                  | 1 n/ms           | Spontaneous fission | $\nu_{S1} = 3.76$ $\sigma = 1.245$       | 12.528  |

In each case, the reference distribution and the distributions reconstructed from the first moments are plotted, which allows for a qualitative assessment of the quality of the reconstruction. For a quantitative evaluation of the distance between the reference and approximated distributions, we also calculate the Euclidean distance *D*.

$$D = \sqrt{\sum_{i=1}^{N}\left(P_i^{\text{Ref}} - P_i^{\text{App}}\right)^2}$$

The Euclidean distances for the four test cases are reported in table III.

**TABLE III.** Euclidean distances between the exact and approximated distributions.

| Distributions      | Moments | Case 1 - *D*    | Case 2 - *D*    | Case 3 - *D*    | Case 4 - *D*    |
|--------------------|---------|-----------------|-----------------|-----------------|-----------------|
| Polyà              | 2       | 0.112 10$^{-1}$ | 0.786 10$^{-2}$ | 0.146           | 0.798 10$^{-1}$ |
| Meixner expansion  | 4       | 0.244 10$^{-2}$ | 0.190 10$^{-2}$ | 0.119           | 0.621 10$^{-1}$ |
| Quadratic (2-forked) | 2     | 0.542 10$^{-3}$ | 0.510 10$^{-2}$ | 0.886 10$^{-2}$ | 0.791 10$^{-1}$ |
| Cubic (3-forked)   | 3       | 0.122 10$^{-4}$ | 0.470 10$^{-3}$ | 0.162 10$^{-2}$ | 0.339 10$^{-1}$ |
| Quartic (4-forked) | 4       | 0.848 10$^{-5}$ | 0.572 10$^{-4}$ | 0.639 10$^{-3}$ | 0.171 10$^{-1}$ |

The distributions obtained in case 1 are presented in figure 2 and 3. Here we are in favorable conditions for the use of the approximations. The detection efficiency is low which leads to a moderate Feynman parameter and the source neutrons are uncorrelated. The distributions shown in figure 2 and 3 appear to be confounded except for the Polya distribution that deviates slightly from the reference. The distances given in table III show that, the quadratic distribution is a better approximation to the reference than the Polyà distribution. The Meixner expansion with four moments improves the Polyà distribution, which remains not as good as the quadratic distribution that uses only two moments. As expected, distributions based on the cubic and quartic approximations improve the quality of the reconstruction.

The distributions corresponding to the second case are plotted in figure 4 and 5. The difference with the first case is the presence of a spontaneous fission source. Because the N-forked branching approach assumes that the source is not of this type, the reconstruction is not as good as in case one but remains very accurate. Polyà's approximation and Meixner's expansion give slightly better results than in case one.

The distributions corresponding to the third case are plotted in figure 6 and 7. Here, the difference with the first case is that the efficiency is higher and the source is lower, the detection rate remains unchanged. This implies a higher level of correlations as shown by the increase in the Feynman parameter $Y_2$. In this case, the N-forked approximations are less accurate than in the previous situations but still satisfactory. On the other hand, the distributions calculated with Polyà's approximation and fourth order Meixner's expansion are not satisfactory, they do not allow finding the shape of the reference distribution.

The distributions corresponding to the fourth case are plotted in figure 8 and 9. The difference with the previous case lies in the type of source that is here a spontaneous fission source. Here we are in a situation less favorable to the application of the approximations studied, which explains the lower quality of the reconstructions. Here again, the best results are obtained with the quartic or 4-forked branching approximation.

So far, for comparison with the other methods, we have limited the order of the expansion on the basis of Meixner polynomials to four. However, the reconstruction can be improved by increasing the number of moments restored. For example, in the third and fourth case, we can recover the shape of the reference distribution (cf. figure 7 and 9). However, it is necessary to go up to order 16 or 32 to exceed the precision of the quadratic approximation as can be seen in table IV.

**TABLE IV.** Euclidean distances between the exact and approximated distributions, using Meixner's polynomial expansions, in case 3 and 4.

| Distributions | Moments | Case 3 - $D$ | Case 4 - $D$ |
|---|---|---|---|
| Polyà | 2 | 0.146 | 0.798 $10^{-1}$ |
| Meixner expansion | 4 | 0.119 | 0.621 $10^{-1}$ |
| Meixner expansion | 8 | 0.680 $10^{-1}$ | 0.345 $10^{-1}$ |
| Meixner expansion | 16 | 0.188 $10^{-1}$ | 0.101 $10^{-1}$ |
| Meixner expansion | 32 | 0.309 $10^{-2}$ | 0.117 $10^{-2}$ |

In general, we find that for the same number of moments used, the N-forked branching approximations are better than the Polyà distribution approximation and the development on the basis of orthogonal Meixner polynomials. Moreover, the accuracy of the reconstruction is excellent when we stay within the conditions of application of the approximations but it degrades when the level of correlations increases.

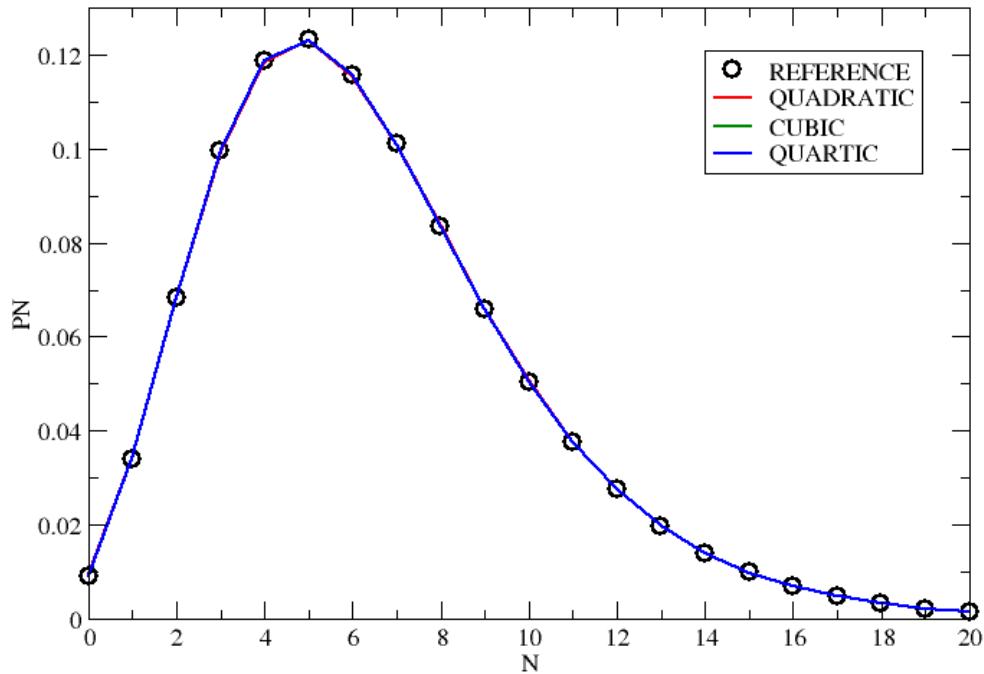

**Figure 2.** Test case 1, Reference, quadratic, cubic and quartic approximated distributions.

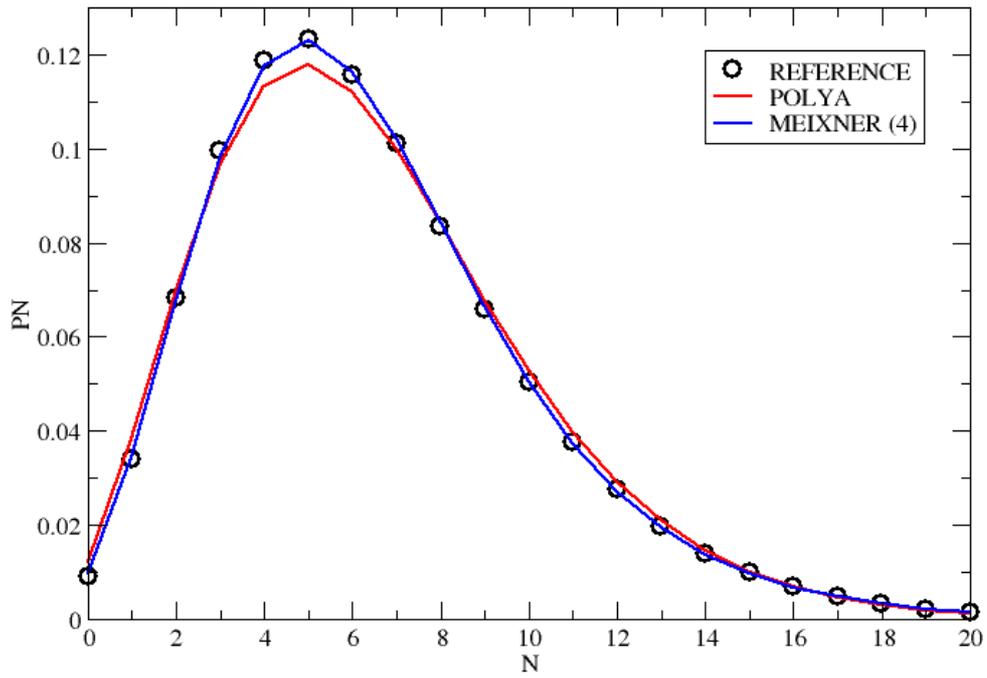

**Figure 3.** Test case 1, Reference, Polyà distributions and Meixner expansion.

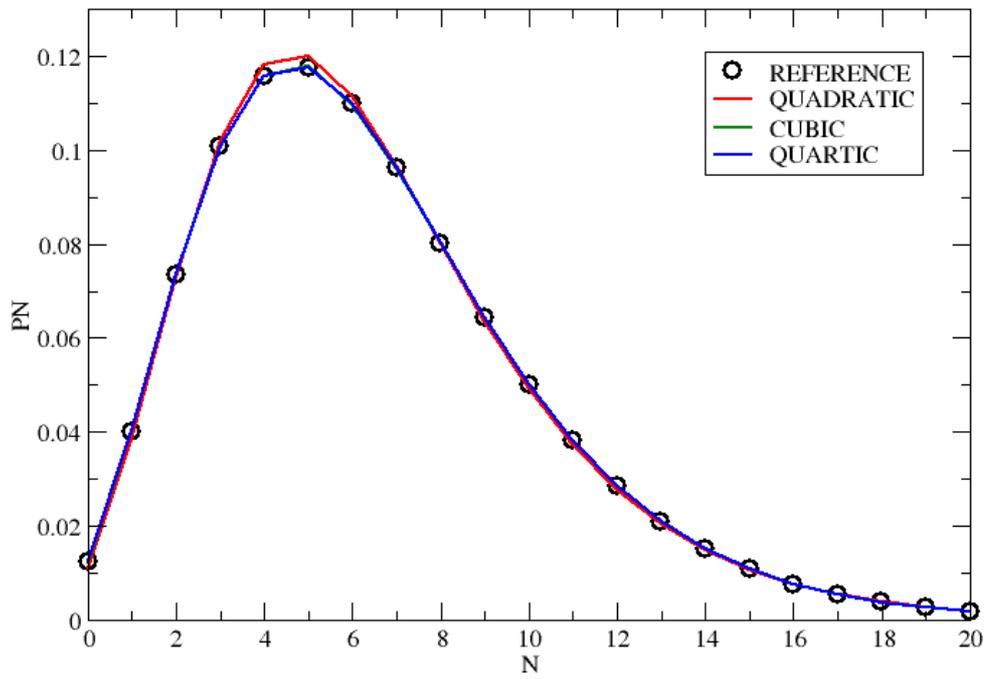

**Figure 4.** Test case 2, Reference, quadratic, cubic and quartic approximated distributions.

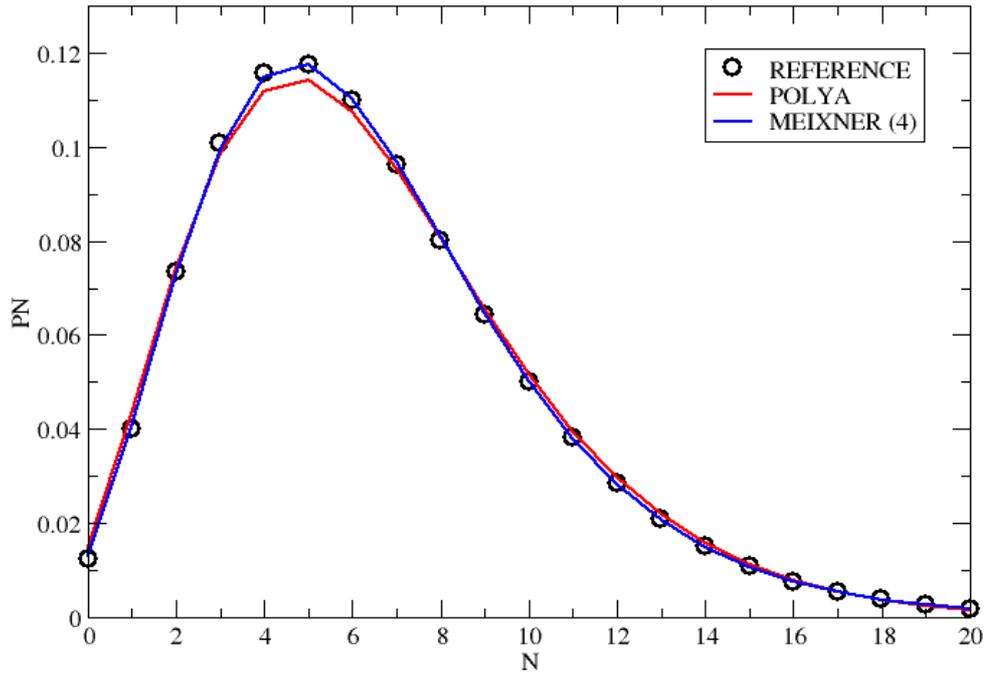

**Figure 5.** Test case 2, Reference, Polyà distributions and Meixner expansion.

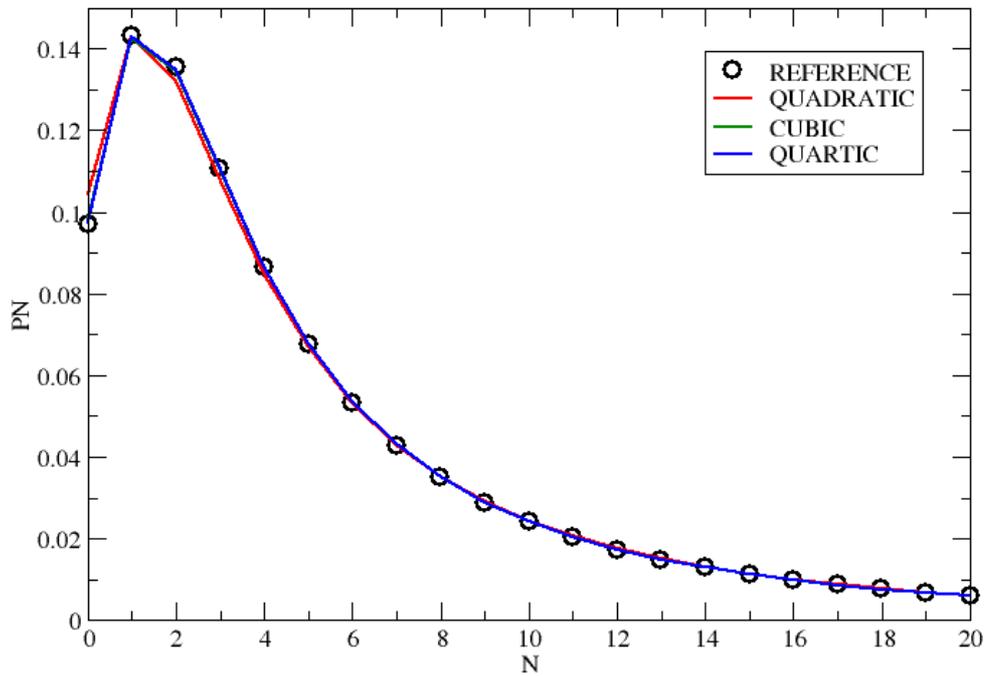

**Figure 6.** Test case 3, Reference, quadratic, cubic and quartic approximated distributions.

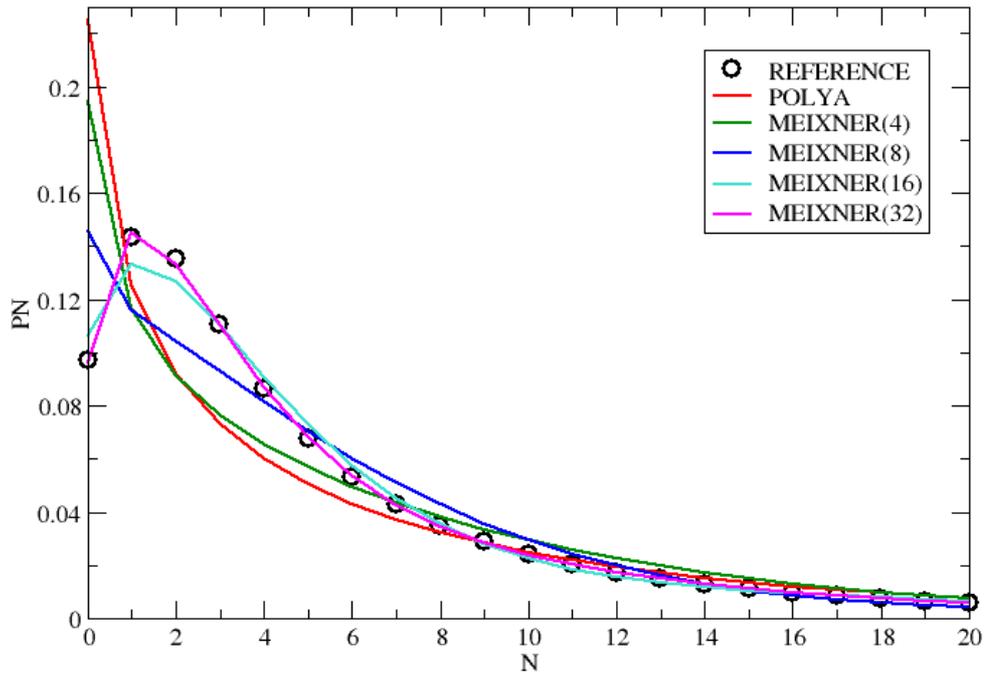

**Figure 7.** Test case 3, Reference, Polyà distributions and Meixner expansions.

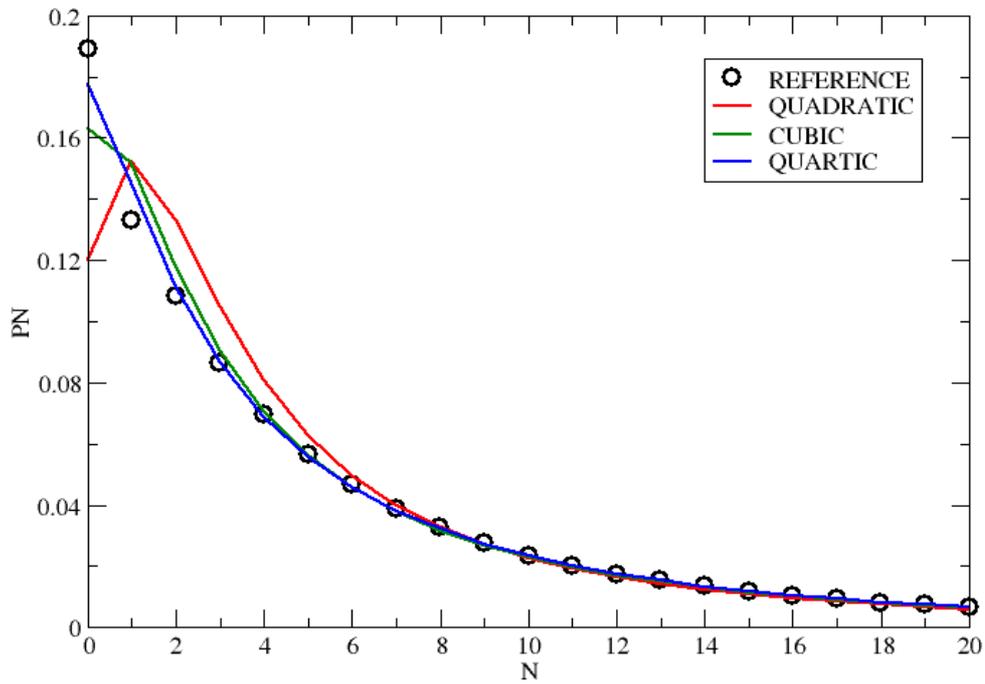

**Figure 8.** Test case 4, Reference, quadratic, cubic and quartic approximated distributions.

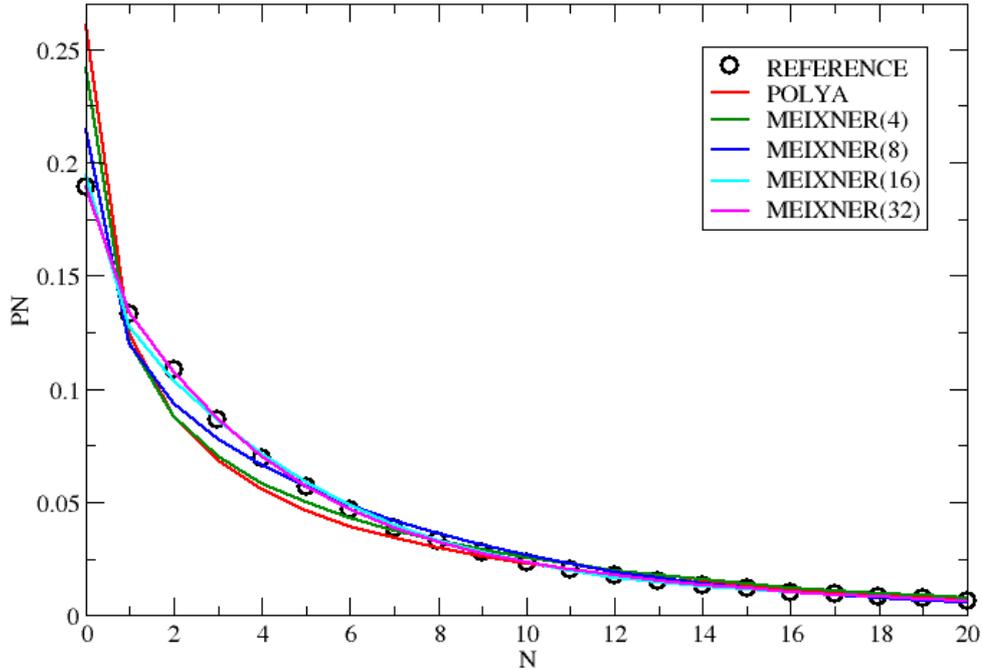

**Figure 9.** Test case 4, Reference, Polyà distributions and Meixner expansions.

**VI. CONCLUSION**

We have developed and implemented two methods for constructing an approximate distribution of the neutron detection number using only the first moments of the original distribution. We have also presented a recursive calculation of the reference distributions in a point model approximation.

The first approach is a generalization of the Poisson Radical distribution that has been extended by taking into account the three and four-forked branching configurations of the fission chains producing the detections. The approximated distributions are parametrized by the first moments. Although they do not have an analytic expression, they can be calculated efficiently using the presented recurrence methods.

The second method is an improvement of the Polyà distribution approximation that uses an expansion of the reference distribution on the basis of Meixner discrete orthogonal polynomials. The first approximation using only two moments is the Polyà distribution.

In general, the reconstructed distributions remain accurate as long as the correlation level of the detected neutrons stays moderate. For a given number of moments, the N-forked branching method gives better results and the accuracy increases with the number of moments. In our test cases, the best approximation with four moments is obtained with the four-forked or quartic approximation.

It is expected that, in most applications of fissile materials detection, the approximate distributions could be accurately and quickly calculated so that they could be used to estimate nuclear parameters from neutron correlation measurements using inverse problem techniques like Bayesian inversion and MCMC sampling methods.

**REFERENCES**

1. R.E. UHRIG, *Random Noise Techniques in Nuclear Reactor Systems*, Ronald Press, New York (1970).
2. I. PAZSIT and L. Pal, *Neutron fluctuation, a treatise on the physics of branching processes*, Elsevier, Oxford (2008).
3. K. BÖHNEL, "The Effect of Multiplication on the Quantitative Determination of Spontaneously Fissioning Isotopes by Neutron Correlation Analysis," *Nuclear Science and Engineering*, **90**, 75-82 (1985); https://doi.org/10.13182/NSE85-2
4. R.P. FEYNMAN, F. DE HOFFMANN and R. SERBER, "Dispersion of the neutron emission in U-235 fission," *Journal of Nuclear Energy,* **3**, 64–69 (1956); https://doi.org/10.1016/0891-3919(56)90042-0
5. M.K. PRASAD and N.J. SNYDERMAN, "Statistical Theory of Fission Chains and Generalized Poisson Neutron Counting Distributions," *Nuclear Science and Engineering*, **172**, 300-326 (1985); dx.doi.org/10.13182/NSE11-86
6. J. VERBEKE, "Neutron multiplicity counting: credible regions for reconstruction parameters," *Nuclear Science and Engineering***,** **182** (4), 481-501 (2016); dx.doi.org/10.13182/NSE15-35
7. L. PÀL, "Statistical Theory of Neutron Chain Reactions I, II ,III," *Acta Physica Hungaria,* **14** (4), pp. 345–369 (1962).
8. V.G. ZOLOTUKHIN and I. MOGILNER, "The Distribution of Neutron Counts from a Detector Placed in a Reactor," *Atomnaya Energiya*, **10** (4), pp.379-381 (1961).
9. G.I. BELL, "Probability Distribution of Neutrons and Precursors in a Multiplying Assembly," *Annals of Physics*, **21** (2), pp. 243–283 (1963).
10. D. BABALA, "Neutron Counting Statistics in Nuclear Reactors," KR-114, Institutt for Atomenergi, Kjeller (1966).
11. N. PACILIO, "The Polyà Model and the Distribution of Neutrons in a Steady State reactor", *Nuclear Science and Engineering*, **26** (4), pp. 565-569 (1966); https://doi.org/10.13182/NSE66-A18430
12. N. PACILIO and V.M. JORIO, "Discrete Probability Distributions in Nuclear Reactor Theory," *Annals of Nuclear Energy*, **2** (2-5), pp. 89-93 (1975); https://doi.org/10.1016/0306-4549(75)90008-0
13. J.R. MOUSSA and A.K. PRINJA, "Reconstruction of Neutron Multiplicity Distributions from Low-Order Statistical Information, " *Nuclear Instruments and Methods in Physics Research - section A*, in Press, (2022); https://doi.org/10.1016/j.nima.2022.167429
14. A.F. NIKIFOROV, S.K. SUSLOV and V.B. UVAROV, *Classical Orthogonal Polynomials of a Discrete Variable*, Springer-Verlag, Berlin Heidelberg (1991).
15. L. PAL, "On the Theory of Stochastic Processes in Nuclear Reactors," *Il Nuovo Cimento*, **7**, pp. 25–42 (1958).
16. G.I. BELL, "On the Stochastic Theory of Neutron Transport," *Nuclear Science and Engineering*, **21**, 390-401 (1965); https://doi.org/10.13182/NSE65-1
17. H.H. PANJER, "Recursive Evaluation of a Family of Compound Distributions," *ASTIN Bulletin*, **12** (1), 22-26 (1981); https://doi.org/10.1017/S0515036100006796
18. J. TERRELL, "Distribution of Fission Neutron Numbers," *Physical review C*, **1**, 783 (1957); https://doi.org/10.1103/PhysRev.108.783